\documentclass[aps,pre,reprint,superscriptaddress,amsmath,amssymb,longbibliography,floatfix]{revtex4-2}
\usepackage[T1]{fontenc}
\usepackage{CJK}
\usepackage{graphicx,bm,booktabs,dcolumn,xcolor}
\usepackage[colorlinks=true,linkcolor=blue!55!black,citecolor=blue!55!black,urlcolor=blue!55!black]{hyperref}
\hypersetup{pdftitle={Initial heat-current decay versus thermal transport at a three-body resonance in a one-dimensional Fermi gas},pdfauthor={Yixin Guo}}
\graphicspath{{figures/}}
\newcommand{\dd}{\mathrm d}
\newcommand{\ii}{\mathrm i}
\newcommand{\e}{\mathrm e}
\newcommand{\He}{\mathrm{He}}
\newcommand{\Ree}{\operatorname{Re}}
\newcommand{\Imm}{\operatorname{Im}}
\newcommand{\tr}{{\rm tr}}
\newcommand{\intk}{\int\!\frac{\dd k}{2\pi}}
\begin{document}
\begin{CJK}{UTF8}{ipxm}
\title{Initial heat-current decay versus thermal transport at a three-body resonance\\
in a one-dimensional Fermi gas}
\author{Yixin Guo (郭一昕)}
\email{yixin.guo@riken.jp}
\affiliation{RIKEN Nishina Center for Accelerator-Based Science, Wako 351-0198, Japan}
\date{October 17, 2026}

\begin{abstract}
We theoretically study kinetic relaxation and heat transport in a dilute one-dimensional spinless Fermi gas near a three-body resonance.
The full vacuum amplitude of a finite-range interaction enters the linearized Boltzmann collision operator.
The resonance position that maximizes the initial heat-current decay rate lies above the one that minimizes the integrated relaxation time.
At higher resonance energies, coupling among momentum moments transfers heat-current weight to a slow mode.
A faster initial decay can therefore coexist with slower subsequent relaxation and a larger low-frequency heat response.
The dissipative responses at two resonance positions reverse their ordering at least twice as frequency increases.
Bounds on the full collision operator confirm both reversals.
A fourth-moment distortion exhibits the same separation of relaxation scales and provides a probe through the momentum distribution.
\end{abstract}
\maketitle
\end{CJK}

\section{Introduction}
\label{sec:intro}

Relaxation in one-dimensional gases is strongly influenced by integrability and its associated conservation laws~\cite{Kinoshita2006,Rigol2007}.
Experiments with split Bose gases have observed relaxation to prethermalized states~\cite{Gring2012}.
In a dipolar quantum Newton's cradle, tunable integrability breaking produces a later stage of thermalization~\cite{Tang2018}.
Kinetic descriptions connect such slow dynamics to collisions that relax the additional conserved quantities~\cite{Bastianello2021}.
For coupled quasi-one-dimensional Bose gases, such a collision integral yields a broad spectrum of time scales and nonexponential relaxation~\cite{Panfil2023}.

The kinematics of one dimension also restricts the redistribution of momenta.
An elastic collision between two equal-mass particles can only preserve or exchange their momenta.
Non-integrable three-particle collisions are diffractive and allow a continuous set of outgoing configurations~\cite{Lamacraft2013}.
In a dilute gas of identical particles, three-body collisions therefore provide the leading relaxation channel.
Virtual transverse excitations generate such effective three-body processes in quasi-one-dimensional Bose gases, where they break integrability and permit thermalization~\cite{Mazets2008,Mazets2010}.

Three-body interactions also shape the few-body spectrum in one dimension.
Two- and three-body attractions produce universal trimer states of bosons~\cite{Nishida2018,Guijarro2018}, and confinement can induce trimers directly~\cite{Pricoupenko2018}.
An effective gas with a pure three-body repulsion has also been proposed using dimers near a dimer-dimer zero crossing~\cite{Guijarro2018}.
Fermion-mediated two- and three-body interactions produce bosonic clusters in a one-dimensional Bose-Fermi mixture~\cite{Guo2024A}.
At low energy, three-body scattering is characterized by a scattering hypervolume~\cite{Tan2008}.
This quantity has also been determined for identical fermions in one dimension~\cite{Wang2023}.

Three-body collisions are known to control thermal transport in several regimes.
The thermal conductivity of a weakly interacting quasi-one-dimensional Bose gas has been calculated from the linearized Boltzmann equation~\cite{Tanaka2022}.
The relaxation-time approximation underestimates the full conductivity by more than 20\% at the lowest temperature considered.
At finite frequency, this approximation gives a Drude response.
Three-particle collisions also govern thermoelectric corrections and hot-carrier relaxation in clean quantum wires~\cite{Lunde2007,Karzig2010}.
Collision spectra resolve the rates associated with equilibration~\cite{Micklitz2011}.
The thermal conductance distinguishes intrabranch and interbranch relaxation~\cite{Levchenko2011}.
Distinct equilibration rates and transport regimes also occur in degenerate one-dimensional Fermi gases~\cite{Matveev2019}.
For resonating two-component fermions in two and three dimensions, the linearized Boltzmann equation has been derived microscopically~\cite{Fujii2021}.
The resulting transport coefficients show small deviations from the relaxation-time approximation.

In earlier work, we studied the competition between pairing and tripling in two-component fermions with coexisting even- and odd-parity interactions~\cite{Guo2023B}.
For spinless fermions, we used the generalized Cooper problem to study three-body clustering above a Fermi sea~\cite{Guo2022}.
We subsequently studied an attractive three-body interaction with the minimal antisymmetric momentum dependence, both alone and together with two-body forces~\cite{Guo2023A}.
The same antisymmetric three-body channel also allows momentum redistribution in the scattering continuum.

A resonance in this channel selects the relative energy at which collisions occur.
In this paper, we study a dilute spinless Fermi gas with a pure three-body interaction of this leading antisymmetric form~\cite{Guo2023A}.
A Gaussian form factor sets the interaction range.
The separable model gives a closed vacuum amplitude, whose full energy dependence enters the collision operator.
We find that the resonance energy that maximizes the initial heat-current decay rate differs from the one that optimizes integrated heat relaxation.
The initial heat-current rate peaks for a resonance near six thermal energies, as fixed by the threshold law of the antisymmetric vertex.
The integrated heat-current relaxation is fastest near four thermal energies.
Between one and four thermal energies, the integrated time is less than twice the inverse initial rate.
Near $u_r=4$, the coupling between the heat current and the next odd moment is weak.
At higher energies, resonant collisions couple several odd moments into a slow heat-current mode.
At the rate maximum, the integrated relaxation time exceeds the inverse initial rate by a factor between 4.6 and 7.7 for the temperatures studied.
This large departure from a relaxation-time description contrasts with resonating fermions in two and three dimensions, where the Prandtl number stays close to its relaxation-time value~\cite{Fujii2021}.
A larger initial heat-current decay rate can therefore produce a larger low-frequency heat response.
We confirm this with bounds on the full collision operator, which also establish two reversals in the ordering of the finite-frequency responses.
A fourth-moment distortion of the momentum distribution, whose initial rate is exactly related to that of the heat current, provides an independent and directly measurable probe of the same physics.

This paper is organized as follows.
In Sec.~\ref{sec:theory}, we introduce the interaction and derive the collision kinematics, collision moments, and response functions.
In Sec.~\ref{sec:results}, we present the numerical results and their physical origin.
A summary and perspectives are given in Sec.~\ref{sec:summary}.
The amplitude normalization, the construction of collision matrices, and the full-operator bounds are given in the Appendixes.
We take $\hbar=k_{\rm B}=1$ throughout this paper.

\section{Theoretical framework}
\label{sec:theory}

\subsection{Pure three-body interaction}

We consider identical spinless fermions of mass $m$ in a homogeneous one-dimensional continuum.
The $N$-particle Hamiltonian is
\begin{equation}
 H_N=\sum_{i=1}^N\frac{k_i^2}{2m}
       +\sum_{i<i'<i''}v_3^{(ii'i'')},
 \label{eq:H}
\end{equation}
and it contains no pairwise interaction.
For a triple, the Jacobi momenta are
\begin{align}
 P&=k_1+k_2+k_3,\\
 p&=(k_1-k_2)/\sqrt2,\qquad
 q=(k_1+k_2-2k_3)/\sqrt6.
\end{align}
The total energy is $P^2/(6m)+E$, with relative energy $E=\rho^2/(2m)$ and $\bm\rho=(p,q)$.
Relative momentum states are normalized as $\langle\bm\rho'|\bm\rho\rangle=(2\pi)^2\delta^{(2)}(\bm\rho'-\bm\rho)$.
We use a rank-one relative interaction,
\begin{align}
 \langle\bm\rho'|v_3|\bm\rho\rangle&=g_3 F(\bm\rho')F(\bm\rho),\\
 F(\bm\rho)&=\frac{p(p^2-3q^2)}{\sqrt2}
             \e^{-\rho^2/(2\Lambda^2)},
 \label{eq:F}
\end{align}
where $g_3$ is the three-body interaction strength.
The momentum cutoff scale $\Lambda$ sets the interaction range.
The polynomial in Eq.~\eqref{eq:F} equals $(k_1-k_2)(k_2-k_3)(k_3-k_1)$ and changes sign under every particle exchange.
It is the leading antisymmetric structure of Ref.~\cite{Guo2023A}.
The Gaussian factor depends only on relative momenta and preserves Galilean invariance.
For spinless fermions with an odd-wave contact interaction, a marginal three-body coupling between a fermion and a $p$-wave dimer is needed to regularize the three-body sector~\cite{Sekino2021}.
The present vertex acts directly on three fermions and contains no pairwise channel.

At leading order in fugacity, the equilibrium single-particle momentum distribution $f_{k,0}$ takes the Boltzmann form
\begin{equation}
 f_{k,0}=z\e^{-k^2/(2mT)},\qquad
 n=z\sqrt{\frac{mT}{2\pi}},
 \label{eq:f0}
\end{equation}
where $z=\e^{\beta\mu}\ll1$ is the fugacity and $n$ is the density.
The subscript $0$ labels the equilibrium state.
Here $\beta=1/T$ is the inverse temperature and $\mu$ is the chemical potential.

\subsection{Vacuum amplitude and resonance parameters}

The vacuum three-body $T$ matrix has the separable relative-momentum amplitude $F(\bm\rho')\tau(E)F(\bm\rho)$.
The scalar amplitude $\tau(E)$ contains the energy dependence from repeated scattering and obeys
\begin{align}
 \tau(E)^{-1}&=g_3^{-1}-\Pi(E),\\
 \Pi(E)&=\frac1{8\pi}\int_0^\infty\dd\rho\,
 \frac{\rho^7\e^{-\rho^2/\Lambda^2}}
 {E+\ii0-\rho^2/(2m)}.
\end{align}
The vacuum loop function $\Pi(E)$ describes free three-particle propagation between successive interactions.
We define the range energy $E_R$, the loop scale $\Pi_0$, and the dimensionless interaction detuning $\eta$ by
\begin{equation}
 E_R=\frac{\Lambda^2}{2m},\qquad
 \Pi_0=\frac{m\Lambda^6}{8\pi},\qquad
 \eta=\frac{g_3^{-1}+2\Pi_0}{\Pi_0}.
\end{equation}
For positive energy, the amplitude has the closed form
\begin{subequations}
\label{eq:ab}
\begin{align}
 \tau(E)&=\frac1{\Pi_0[a(x)+\ii b(x)]},\qquad x=E/E_R,\\
 a(x)&=\eta+x+x^2-x^3\e^{-x}\operatorname{Ei}(x),\\
 b(x)&=\pi x^3\e^{-x}.
\end{align}
\end{subequations}
Here $a(x)$ and $b(x)$ are the real and imaginary parts of the scaled inverse amplitude $\tau(E)^{-1}/\Pi_0$.
The function $\operatorname{Ei}$ is the exponential integral.
The imaginary part follows from the open elastic channel.
The angular dependence of $F$ selects the interacting scattering channel.
Its elastic $S$-matrix eigenvalue $S_3(E)$ is
\begin{equation}
 S_3(E)=\frac{a(x)-\ii b(x)}{a(x)+\ii b(x)}
       =\e^{2\ii\delta_3(E)},
 \label{eq:S}
\end{equation}
where $\delta_3(E)$ is the elastic three-body scattering phase shift.
Thus $|S_3|=1$, $\Imm\tau=-\Pi_0 b|\tau|^2$, and $\sin^2\delta_3=b^2/(a^2+b^2)$.

Throughout, we write $a'(x)=\dd a(x)/\dd x$ for the derivative at fixed detuning $\eta$.
We tune the low-energy phase crossing $a(x_r)=0$ with $a'(x_r)>0$ and denote its energy by $E_r=x_rE_R$.
We use the reduced temperature $\vartheta$, the scaled collision energy $u$, and the scaled resonance energy $u_r$,
\begin{equation}
 \vartheta=T/E_R,\qquad u=E/T,\qquad u_r=E_r/T.
\end{equation}
The resonance position therefore satisfies $x_r=\vartheta u_r$.
The linearized resonance width $\Gamma_{\rm lin}$ and the Wigner time delay $t_W$ at the crossing are given, respectively, by
\begin{equation}
 \Gamma_{\rm lin}=\frac{2E_Rb(x_r)}{a'(x_r)},\qquad
 t_W=\frac{2a'(x_r)}{E_Rb(x_r)}
           =\frac4{\Gamma_{\rm lin}}.
 \label{eq:delay}
\end{equation}
The width is the full width at half maximum of the linearized resonance profile.
The Wigner delay measures the delay relative to free propagation and equals twice the energy derivative of the phase shift at $E_r$.
At zero energy, the fermionic three-body scattering hypervolume $D_F$ in the convention of Ref.~\cite{Wang2023} is
\begin{equation}
 D_F=72\sqrt3\,m\tau(0)
    =\frac{576\sqrt3\pi}{\Lambda^6\eta},
 \label{eq:hypervolume}
\end{equation}
which has dimensions of length to the sixth power and diverges at the threshold condition $\eta=0$.
The matching and the physical Slater amplitude are derived in Appendix~\ref{app:norm}.

\subsection{Collision operator and heat response}

Let $f_k(t)$ be the single-particle momentum distribution at time $t$, with initial value $f_k(0)$.
We write its small departure from equilibrium as $\delta f_k(t)=f_k(t)-f_{k,0}=f_{k,0}\phi(k,t)$.
The dimensionless function $\phi(k,t)$ describes the relative distortion of the distribution.
The linearized three-body collision operator $\mathcal L$ generates the homogeneous evolution $\partial_t\phi=-\mathcal L\phi$.
An eigenmode satisfying $\mathcal L\phi_\lambda=\lambda\phi_\lambda$ decays as $\e^{-\lambda t}$.
The eigenvalue $\lambda$ is its relaxation rate.
Let $\int_{1\cdots6}$ denote six momentum integrals with measure $\dd k/(2\pi)$, with incoming labels $123$ and outgoing labels $456$.
The antisymmetrized three-fermion scattering amplitude $M_{456;123}$ and its transition kernel $W_{123\to456}$ are
\begin{align}
 M_{456;123}&=6\sqrt3\,F(\bm\rho_{\rm out})\tau(E)F(\bm\rho_{\rm in}),\label{eq:M}\\
 W_{123\to456}&=(2\pi)^2|M_{456;123}|^2
          \delta(P_{\rm in}-P_{\rm out})\delta(E_{\rm in}-E_{\rm out}),
 \label{eq:W}
\end{align}
where $E_{\rm in}$ and $E_{\rm out}$ are the total kinetic energies.
We abbreviate $f_{i,0}=f_{k_i,0}$ and $\phi_i=\phi(k_i)$.
The incoming minus outgoing change is $\Delta\phi=\phi_1+\phi_2+\phi_3-\phi_4-\phi_5-\phi_6$.
The collision operator is
\begin{equation}
 (\mathcal L\phi)(k_1)=\frac1{12}\int_{2\cdots6}
 W_{123\to456}f_{2,0}f_{3,0}\Delta\phi .
\end{equation}
We hold $k_1$ fixed and integrate over the other five momenta.
The factor $1/12$ counts unordered incoming partners and outgoing triples.
We define the normalized inner product $(\phi,\phi')_0$ with equilibrium weight $f_{k,0}/n$ by
\begin{equation}
 (\phi,\phi')_0=\frac1n\intk f_{k,0}\phi^*(k)\phi'(k).
\end{equation}
The superscript $*$ denotes complex conjugation.
The normalization gives $(1,1)_0=1$.
Detailed balance then gives the positive quadratic form
\begin{equation}
 (\phi,\mathcal L\phi)_0=
 \frac1{72n}\int_{1\cdots6}W_{123\to456}
 f_{1,0}f_{2,0}f_{3,0}|\Delta\phi|^2.
 \label{eq:quadratic}
\end{equation}
Thus $\mathcal L$ is self-adjoint and nonnegative.
The functions $1$, $k$, and $k^2$ are collision invariants associated with conservation of particle number, momentum, and energy, respectively.
For example, $\phi(k)=1$ counts each particle once.
Each three-body collision therefore has $\Delta\phi=3-3=0$, giving $\mathcal L1=0$.
We will show in Sec.~\ref{sec:kinematics} that these are the only collision invariants for $g_3\ne0$.

We use the thermal momentum variable $y=k/\sqrt{mT}$ and the normalized Hermite basis
\begin{equation}
 \psi_j(y)=\frac{\He_j(y)}{\sqrt{j!}},\qquad
 (\psi_j,\psi_{j'})_0=\delta_{jj'}.
\end{equation}
Here $\He_j$ is the Hermite polynomial of order $j$ orthogonal under the Gaussian weight $\e^{-y^2/2}$.
The modes $\psi_0$, $\psi_1$, and $\psi_2$ span the conserved space, and all matrices below act on its complement.
The collision matrix $\mathbf L$ has the elements
\begin{equation}
 L_{jj'}=(\psi_j,\mathcal L\psi_{j'})_0,\qquad j,j'\ge3,
\end{equation}
and momentum inversion separates its odd and even sectors.

The heat current carried by a particle in the rest frame is
\begin{equation}
 Q(k)=\frac{k}{m}\left(\frac{k^2}{2m}-\frac{3T}{2}\right)
 =\sqrt{\frac32}\frac{T^{3/2}}{\sqrt m}\psi_3(y).
 \label{eq:heat}
\end{equation}
Subtracting the enthalpy per particle $3T/2$ removes the convective contribution.
Linear-response theory relates thermal transport to equilibrium heat-current correlations~\cite{Kubo1957,Luttinger1964}.
The corresponding kinetic thermal conductivity is
\begin{equation}
 \kappa(\omega)=
 \frac{3nT}{2m}\,(\psi_3,(\mathcal L-\ii\omega)^{-1}\psi_3)_0
 \label{eq:kappa}
\end{equation}
where $\omega>0$ is the angular frequency of the thermal perturbation.
The retarded prescription selects the causal response, and $\Ree\kappa(\omega)$ is its dissipative part.
Its prefactor does not depend on the detuning at this order.

We define the normalized autocorrelation function $C_a(t)$ of a mode $\psi_a$ by
\begin{equation}
 C_a(t)=(\psi_a,\e^{-\mathcal Lt}\psi_a)_0
 =\int_0^\infty\e^{-\lambda t}\dd\nu_a(\lambda).
 \label{eq:Ca}
\end{equation}
For $a=3$, $C_3(t)$ is the normalized heat-current correlation.
For $a=4$, the mode $\psi_4$ describes a distortion of the fourth moment of the momentum distribution.
This distortion preserves particle number, momentum, and energy.
We will study its relaxation in Sec.~\ref{sec:fourth}.
The spectral measure $\nu_a$ gives the distribution of source weight over relaxation rates and is normalized to unity.
For each source, the initial decay rate is the mean of its spectral measure,
\begin{equation}
 \gamma_a=-\left.\frac{\dd C_a(t)}{\dd t}\right|_{t=0^+}
 =\int\lambda\,\dd\nu_a(\lambda)=L_{aa},
 \label{eq:initial}
\end{equation}
For the heat-current source $a=3$, we write $\gamma_Q\equiv\gamma_3=L_{33}$, with $Q$ referring to the heat current in Eq.~\eqref{eq:heat}.
The time integral of $C_3(t)$ defines the heat transport rate,
\begin{equation}
 \gamma_{\tr}^{-1}=\int_0^\infty C_3(t)\,\dd t
 =\int\lambda^{-1}\dd\nu_3(\lambda),
 \label{eq:transport}
\end{equation}
which fixes the low-frequency limit $\kappa(\omega\to0)=3nT/(2m\gamma_\tr)$ of Eq.~\eqref{eq:kappa}.
The heat-current memory factor
\begin{equation}
 K=\frac{\gamma_Q}{\gamma_\tr}
 =\int\lambda\,\dd\nu_3\int\lambda^{-1}\dd\nu_3\ge1
 \label{eq:K}
\end{equation}
equals unity only for a single relaxation rate.
The time $t$ refers to times long compared with the collision delay, as discussed in Sec.~\ref{sec:validity}.

\subsection{Collision kinematics}
\label{sec:kinematics}

We decompose the dimensionless momenta of a colliding triple as $y_i=s+\xi_i$.
Here $s=P/(3\sqrt{mT})$ is the scaled center-of-mass momentum per particle, and $\xi_i$ are relative momenta satisfying $\sum_i\xi_i=0$.
We define the power sums $\Sigma_j=\sum_{i=1}^3(\xi_i)^j$, where $j$ denotes the power of each relative momentum.
Momentum and energy conservation fix $s$ and $\Sigma_2=2u$ in a collision at relative energy $uT$.
For three numbers with zero sum, $\Sigma_4=(\Sigma_2)^2/2$ and $\Sigma_5=5\Sigma_2\Sigma_3/6$.
Here $(\Sigma_2)^2$ denotes the square of the sum of squared relative momenta.
At fixed $\Sigma_2$, all other power sums are polynomials in the cubic sum $\Sigma_3=3\xi_1\xi_2\xi_3$.
This sum changes with the relative hyperangle, the polar angle in the Jacobi momentum plane $(p,q)$.

These relations show that $1$, $y$, and $y^2$ are the only collision invariants.
For a single collision, $\Delta$ denotes the incoming value minus the outgoing value.
For example, $\Delta\Sigma_3=\sum_{i=1}^3(\xi_{i,\rm in})^3-\sum_{i=1}^3(\xi_{i,\rm out})^3$ compares the cubic relative-momentum sums of the two triples.
For an invariant $\phi(y)$, the sum $\sum_i\phi(s+\xi_i)$ must be unchanged by every collision.
For a smooth $\phi$ at fixed $s$ and $u$, a Taylor expansion about $y=s$ gives a change $\frac16\frac{\dd^3\phi(s)}{\dd s^3}\Delta\Sigma_3+O(u^2)$ in this sum.
Terms through second order cancel because $\Sigma_1=0$ and $\Sigma_2$ is fixed.
The amplitude connects almost every pair of hyperangles at every relative energy, and pairs with different $\Sigma_3$ give $|\Delta\Sigma_3|\propto u^{3/2}$.
Letting $u\to0$ gives $\dd^3\phi(s)/\dd s^3=0$ for every $s$.
Because this condition is invariant under translations of $s$, convolving a square-integrable invariant with a smooth kernel reduces it to the smooth case.

For the low moments, $\Delta$ acts on the entire three-particle Hermite sum.
The same relations give
\begin{subequations}
\label{eq:shell}
\begin{align}
 \Delta\sum_i\He_3(y_i)&=\Delta\Sigma_3,\\
 \Delta\sum_i\He_4(y_i)&=4s\,\Delta\Sigma_3,\\
 \Delta\sum_i\He_5(y_i)&=\left(10s^2+\frac{5u}{3}-10\right)\Delta\Sigma_3.
\end{align}
\end{subequations}
The equilibrium centers have $\langle s^2\rangle=1/3$.
Averaged over the centers, the fifth-moment prefactor is therefore $5(u-4)/3$.
A collision at $u=4$ removes heat current without creating, on average, a fifth-moment distortion.

\subsection{Collision moments}

The center-of-mass and angular integrals in Eq.~\eqref{eq:quadratic} can be performed analytically.
The relative-energy weight $w(u)$ for heat-current decay and its normalization $Z$ are
\begin{equation}
 w(u)=\frac{u^9\e^{-(1+2\vartheta)u}}
                 {a(\vartheta u)^2+b(\vartheta u)^2},
 \qquad Z=\int_0^\infty\dd u\,w(u).
 \label{eq:w}
\end{equation}
We define the normalized energy average of a function $h(u)$ by
\begin{equation}
 \langle h(u)\rangle_w=\frac1Z\int_0^\infty\dd u\,w(u)h(u).
 \label{eq:energyaverage}
\end{equation}
Every polynomial collision element has the form
\begin{equation}
 L_{jj'}=\gamma_Q\langle G_{jj'}(u)\rangle_w,\qquad
 \frac{\gamma_Q}{T}=\frac{\sqrt3\pi}{9}z^2\vartheta^6 Z,
 \label{eq:matrix}
\end{equation}
where the dimensionless polynomials $G_{jj'}(u)$ describe coupling between moment modes $j$ and $j'$ at fixed collision energy $uT$.
Their positive construction is given in Appendix~\ref{app:gram}.
The first odd block is
\begin{equation}
 \begin{pmatrix}G_{33}&G_{35}\\ G_{53}&G_{55}\end{pmatrix}
 =\begin{pmatrix}
 1&\dfrac{\sqrt5}{6}(u-4)\\[3pt]
 \dfrac{\sqrt5}{6}(u-4)&\dfrac5{36}(u^2-8u+24)
 \end{pmatrix}.
 \label{eq:oddblock}
\end{equation}
The coupling $G_{35}$ is proportional to the center average found in Eq.~\eqref{eq:shell}.
With the mean $\bar u=\langle u\rangle_w$ and variance $\sigma_u^2=\langle(u-\bar u)^2\rangle_w$ of the collision energy, the first coupling is
\begin{equation}
 L_{35}/\gamma_Q=\frac{\sqrt5}{6}(\bar u-4).
 \label{eq:coupling}
\end{equation}

The power $u^9$ in Eq.~\eqref{eq:w} has a simple origin.
The squared antisymmetric vertices supply six powers of relative energy, and the squared heat-current change supplies three more.
For a constant low-energy amplitude $\tau_0$ and a negligible range factor, $\vartheta u\ll1$, the initial rate is
\begin{equation}
 \gamma_Q^{\rm far}=
 80640\sqrt3\,n^2m^7|\tau_0|^2T^6 .
 \label{eq:far}
\end{equation}
At the threshold $\eta=0$, the amplitude is $\tau(E)\simeq E_R/(\Pi_0E)$, and the rate becomes the reference rate
\begin{equation}
 \frac{\gamma_0}{T}=560\sqrt3\pi z^2\vartheta^4.
 \label{eq:gamma0}
\end{equation}
At fixed density, the temperature powers are thus $T^6$ away from resonance and $T^4$ at threshold.
These limits require the amplitude approximations to hold over the weighted energy window.
An isolated narrow resonance removes three powers through its width, so that its contribution $\gamma_Q^{\rm res}$ to the initial heat-current decay rate is
\begin{equation}
 \frac{\gamma_Q^{\rm res}}{\gamma_0}\simeq
 \frac{u_r^6\e^{-(1+\vartheta)u_r}}
 {7!\,\vartheta^2 a'(x_r)}.
 \label{eq:narrow}
\end{equation}
For $\vartheta u_r\ll1$, this gives the thermal weight $u_r^6\e^{-u_r}$, with its maximum at $u_r=6$.
The rate of collision events lacks the factor $u^3$ from the heat-current change, so its narrow-resonance weight $u_r^3\e^{-u_r}$ peaks at $u_r=3$.
All numerical calculations below retain the complete amplitude in Eq.~\eqref{eq:ab}.

\subsection{Mode coupling and memory}
\label{sec:memory}

To see how coupling to other moments slows the relaxation, we separate a source from the other modes in its parity sector,
\begin{equation}
 \mathbf L=\begin{pmatrix}\gamma_a&\mathbf r^\top\\ \mathbf r&\mathbf L_\perp\end{pmatrix}.
\end{equation}
Here $\gamma_a=L_{aa}$ is the initial decay rate of the selected mode $\psi_a$.
The vector $\mathbf r$ contains its couplings to the other nonconserved modes of the same parity.
The matrix $\mathbf L_\perp$ contains the collision elements among those modes, and $\top$ denotes transpose.
For the heat current, $a=3$, $\gamma_a=\gamma_Q$, and $\mathbf r=(L_{53},L_{73},\ldots)^\top$.
The corresponding $\mathbf L_\perp$ has indices $5,7,\ldots$, with elements such as $L_{55}$, $L_{57}$, and $L_{77}$.
Eliminating the other coefficients gives the projected memory equation~\cite{Mori1965}
\begin{equation}
 \frac{\dd C_a(t)}{\dd t}=-\gamma_a C_a(t)
 +\int_0^t\dd t'\,\mathbf r^\top\e^{-\mathbf L_\perp(t-t')}\mathbf r\,C_a(t').
 \label{eq:memory}
\end{equation}
The feedback term describes transfer to other moments and the return to the source.
It leaves the initial slope $-\gamma_a$ unchanged but controls the later decay.
For the heat current, the time integral of Eq.~\eqref{eq:memory} gives
\begin{equation}
 K=\left(1-\frac{\mathbf r^\top\mathbf L_\perp^{-1}\mathbf r}{\gamma_Q}\right)^{-1}.
 \label{eq:schur}
\end{equation}
Let $\mathbf L^{(N)}$ denote the collision matrix restricted to the $N$ odd modes $\psi_3,\psi_5,\ldots,\psi_{2N+1}$.
We define its memory factor by $K^{(N)}=\gamma_Q[(\mathbf L^{(N)})^{-1}]_{33}$, where the indices retain the Hermite orders.
The superscript $(N)$ counts the retained modes.
For $N=2$, the basis is $\{\psi_3,\psi_5\}$, and Eqs.~\eqref{eq:oddblock} and \eqref{eq:schur} give
\begin{equation}
 \begin{aligned}
 K^{(2)}&=\left(1-\frac{L_{35}^2}{L_{33}L_{55}}\right)^{-1}\\
 &=1+\frac{(\bar u-4)^2}{\sigma_u^2+8}.
 \end{aligned}
 \label{eq:K2}
\end{equation}
The three-mode factor $K^{(3)}$ retains $\psi_3$, $\psi_5$, and $\psi_7$.
Its evaluation requires the first four moments of $u$, with the polynomials given in Appendix~\ref{app:gram}.

The function $\e^{-\lambda t}$ is convex in $\lambda$ for $t\ge0$.
Using Eqs.~\eqref{eq:Ca} and \eqref{eq:initial}, Jensen's inequality therefore gives
\begin{equation}
 \begin{aligned}
 C_a(t)&=\int_0^\infty\e^{-\lambda t}\dd\nu_a(\lambda)\\
 &\ge\exp\left[-t\int_0^\infty\lambda\dd\nu_a(\lambda)\right]
 =\e^{-\gamma_at}.
 \end{aligned}
 \label{eq:jensen}
\end{equation}
Thus a single exponential with the same initial rate is a lower bound on the correlation.

\subsection{Fourth-moment distortion}
\label{sec:fourth}

The first independent even source is
\begin{equation}
 \psi_4(y)=\frac{y^4-6y^2+3}{\sqrt{24}}.
 \label{eq:psi4}
\end{equation}
This mode obeys $(1,\psi_4)_0=(y,\psi_4)_0=(y^2,\psi_4)_0=0$.
A linearized distortion proportional to $f_{k,0}\psi_4(y)$ therefore preserves particle number, momentum, and energy.
For the initial deviation $\delta f_k(0)=A f_{k,0}\psi_4(y)$ with a small amplitude $A$, the excess fourth moment is
\begin{equation}
 \langle y^4\rangle_t-3=\sqrt{24}\,A C_4(t).
\end{equation}
We define the average of a function $h(y)$ over the time-dependent momentum distribution by
\begin{equation}
 \langle h(y)\rangle_t=\frac1n\intk f_k(t)h\!\left(\frac{k}{\sqrt{mT}}\right).
 \label{eq:timeaverage}
\end{equation}
The density $n=\intk f_k(t)$ is conserved.
This observable describes a change of kurtosis at fixed variance, which can be read off from momentum distributions.

Equation~\eqref{eq:shell} gives $\Delta\sum_i\psi_4(y_i)=2s\,\Delta\sum_i\psi_3(y_i)$ for every collision.
Since the amplitude is independent of $s$, Eq.~\eqref{eq:quadratic} then gives
\begin{equation}
 \gamma_4=\frac43\gamma_Q.
 \label{eq:ratio}
\end{equation}
The derivation uses only Galilean invariance, elastic three-body kinematics, and the Boltzmann center distribution.
The relation therefore holds for any Galilean-invariant three-body kernel of identical particles in the nondegenerate limit.
It fixes only the initial slopes.
The first even elements are
\begin{subequations}
\label{eq:evenblock}
\begin{align}
 G_{44}(u)&=\frac43,\qquad
 G_{46}(u)=\frac{\sqrt{30}}9(u-4),\\
 G_{66}(u)&=\frac5{18}u^2-\frac{20}9u+\frac{140}{27}
                         +\frac{u^3}{3240}.
\end{align}
\end{subequations}
The cubic term in $G_{66}$ arises because the sixth moment also depends on $(\Sigma_3)^2$.

At fixed particle number and energy, the Boltzmann entropy deficit per particle relative to equilibrium is
\begin{equation}
 \Delta S(t)=\frac1n\intk f_k(t)\ln\frac{f_k(t)}{f_{k,0}}.
 \label{eq:relativeentropy}
\end{equation}
This relative entropy is nonnegative and vanishes at equilibrium.
Its expansion in $\phi(k,t)=\delta f_k(t)/f_{k,0}$ starts at quadratic order.
We denote this term by $\Delta S^{(2)}(t)$, where $(2)$ labels the order in the perturbation.
For an initial distortion $\phi(k,0)=A\psi_a(y)$, the evolution $\phi(t)=A\e^{-\mathcal Lt}\psi_a$ gives
\begin{equation}
 \Delta S^{(2)}(t)=\frac12(\phi(t),\phi(t))_0
                 =\frac{A^2}{2}C_a(2t).
 \label{eq:entropy}
\end{equation}
The correlation $C_4$ thus also measures the entropy recovery of a specified distribution distortion.

\section{Results and discussion}
\label{sec:results}

\subsection{Parameters and numerical method}

We keep the Gaussian range fixed and use $z=0.03$ and
\begin{equation}
 \vartheta=0.003,\ 0.006,\ 0.01,\ 0.02.
\end{equation}
At each temperature, $u_r$ runs from zero to 12 in steps of 0.25, which gives 196 states in total.
At fixed temperature, the density is the same for all values of $u_r$.
Since the fugacity is held fixed, the density changes with temperature.
The normalized collision matrices are independent of $z$ at this order, and their dimensional rates scale as $z^2$.

We retain 96 odd Hermite basis functions, $\psi_3,\psi_5,\ldots,\psi_{193}$.
The corresponding collision matrix has dimension $96\times96$, with $j,j'\in\{3,5,\ldots,193\}$.
The integrated relaxation rates and memory factors below are evaluated with this basis.
The energy integration resolves the finite resonance width and retains the complete nonresonant part.
The matrix elements use the positive Fourier and center-of-mass construction of Appendix~\ref{app:gram}.
The largest relative change of the complex response under the final quadrature refinement is $2.42\times10^{-8}$.
The largest sampled 80-to-96-mode change of the finite-frequency response is $1.20\times10^{-5}$.
For $\gamma_\tr$, the 80-to-96-mode relative change is below $1.1\times10^{-6}$ for all 196 states.

The Galerkin approximation projects the kinetic equation onto the retained Hermite modes.
Let $t_{\rm int}^{(N)}$ denote the time integral in Eq.~\eqref{eq:transport} evaluated with $\mathbf L^{(N)}$.
For a trial function $h$ in $V_N=\operatorname{span}\{\psi_3,\psi_5,\ldots,\psi_{2N+1}\}$, the Boltzmann variational principle~\cite{Jung2007} gives
\begin{equation}
 \begin{aligned}
 t_{\rm int}^{(N)}&=[(\mathbf L^{(N)})^{-1}]_{33}\\
 &=\max_{\substack{h\in V_N\\h\ne0}}
 \frac{|(\psi_3,h)_0|^2}{(h,\mathcal Lh)_0}.
 \end{aligned}
 \label{eq:galerkintime}
\end{equation}
Increasing $N$ enlarges $V_N$, so this maximum cannot decrease.
The initial rate $\gamma_Q=L_{33}$ is unchanged as the basis grows.
Since $K^{(N)}=\gamma_Q t_{\rm int}^{(N)}$, the 96-mode values of $K$ are lower bounds up to quadrature errors.
At nonzero frequency, the residual construction of Appendix~\ref{app:bounds} encloses the response of the full collision operator.
At $\hat\omega=0.1$, the enclosures contain the 96-mode responses at $u_r=3.75$ and 5.75 and are narrower than $\pm2\%$ and $\pm3\%$, respectively.

For comparisons at a common physical frequency, we use the dimensionless frequency $\hat\omega$ and thermal conductivity $\hat\kappa$,
\begin{equation}
 \hat\omega=\omega/\gamma_0,\qquad
 \hat\kappa=\frac{2m\gamma_0}{3nT}\,\kappa.
 \label{eq:normalization}
\end{equation}
At fixed temperature, the reference $\gamma_0$ of Eq.~\eqref{eq:gamma0} is independent of detuning.

\begin{figure*}[t!]
 \includegraphics[width=\textwidth]{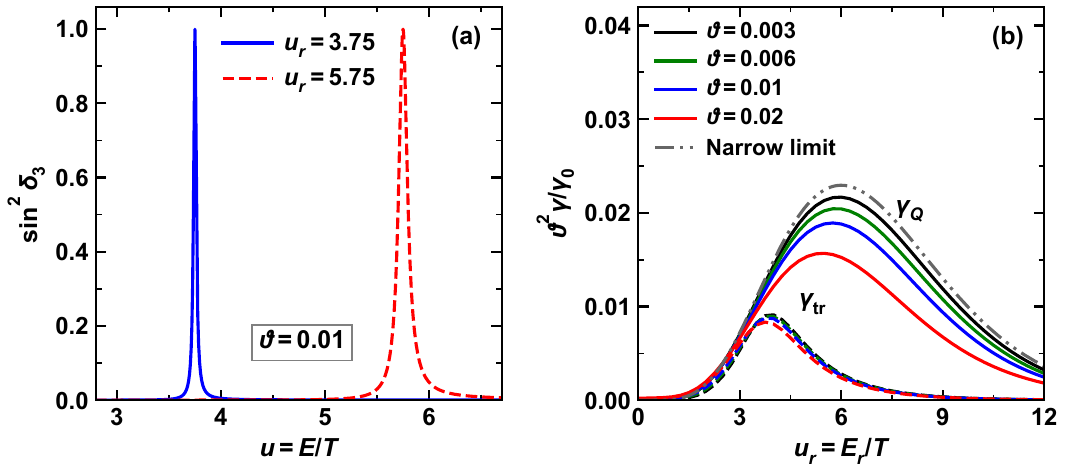}
 \caption{Three-body resonance and the two characteristic rates.
 (a) The elastic probability $\sin^2\delta_3$ at $\vartheta=0.01$ for $u_r=3.75$ and 5.75.
 (b) The initial heat-current decay rate $\gamma_Q$ (solid) and the transport rate $\gamma_\tr$ (dashed), scaled as $\vartheta^2\gamma/\gamma_0$, at four temperatures.
 The transport rates use the 96-mode matrices.
 The gray curve is the narrow-resonance contribution $u_r^6\e^{-u_r}/7!$ to $\gamma_Q$.
 All other curves retain the complete finite-range amplitude.}
 \label{fig:rate}
\end{figure*}

\subsection{Two characteristic resonance energies}

Figure~\ref{fig:rate}(a) shows the elastic probability at two representative detunings.
It reaches unity in the interacting eigenchannel.
The linearized resonance width is given by Eq.~\eqref{eq:delay}.
The thermal rate in Eq.~\eqref{eq:w} additionally weights the energy dependence of the vertices and the heat-current change.

Figure~\ref{fig:rate}(b) compares the two rates of Eqs.~\eqref{eq:initial} and \eqref{eq:transport}.
The initial rate $\gamma_Q$ approaches the narrow-resonance expression as $\vartheta$ decreases.
On the grid of step 0.25, its maximum lies at $u_r=6$, 5.75, 5.75, and 5.5 in increasing order of $\vartheta$.
The resonance slope $a'(x_r)$ in Eq.~\eqref{eq:narrow} depends on the resonance energy.
This dependence and the factor $\e^{-\vartheta u_r}$ shift the maximum at larger $\vartheta$.
The transport rate $\gamma_\tr$ peaks at a clearly lower energy, $u_r=4$, 4, 3.75, and 3.75.
Its scaled maximum, $\vartheta^2\gamma_\tr/\gamma_0=0.0091$, 0.0089, 0.0088, and 0.0083, depends only weakly on temperature.
The resonance energy that maximizes the initial heat-current decay rate therefore differs from the one that makes integrated heat relaxation fastest.
At $\vartheta=0.01$, moving the resonance from $u_r=3.75$ to 5.75 increases $\gamma_Q$ from $117.8\gamma_0$ to $189.2\gamma_0$ but decreases $\gamma_\tr$ from $87.8\gamma_0$ to $31.3\gamma_0$.

\begin{figure*}[t!]
 \includegraphics[width=\textwidth]{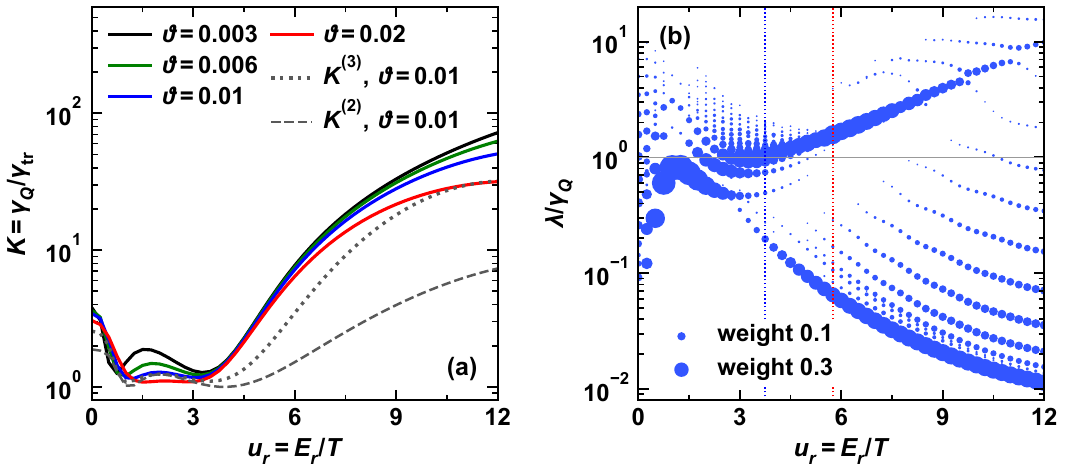}
 \caption{Memory and redistribution of heat-current weight.
 (a) The memory factor $K=\gamma_Q/\gamma_\tr$ at four temperatures.
 The temperature curves use 96 modes.
 Gray curves show the two- and three-mode truncations at $\vartheta=0.01$.
 (b) Eigenvalues $\lambda$ of the 96-mode heat-current matrix at $\vartheta=0.01$.
 The area of each data point is proportional to its spectral weight in $\nu_3$.
 The circles labeled ``weight 0.1'' and ``weight 0.3'' provide reference areas for these weights.
 Data points with weight below $3\times10^{-3}$ are omitted.
 Dotted vertical lines mark $u_r=3.75$ and 5.75, and the horizontal line marks $\lambda=\gamma_Q$.}
 \label{fig:memory}
\end{figure*}

\subsection{Memory factor and slow heat-current mode}
\label{sec:slow}

Figure~\ref{fig:memory}(a) shows how the memory factor of Eq.~\eqref{eq:K} varies with the resonance position $u_r$.
At the threshold $u_r=0$, $K=3.1$--3.8 across the four temperatures.
For $1\le u_r\le4$, $K$ stays below 2 at all four temperatures.
The integrated relaxation time is therefore less than twice the inverse initial rate.
Beyond $u_r\approx4$, $K$ grows rapidly.
At $\vartheta=0.01$, it increases from 1.53 at $u_r=4$ to 6.04 at $u_r=5.75$ and 36.4 at $u_r=10$.
Since $\gamma_Q$ still increases up to $u_r\approx6$, the transport rate $\gamma_\tr=\gamma_Q/K$ peaks near the onset of this rapid growth.
For $4\lesssim u_r\lesssim7$, $K$ grows by a factor of about 7 to 9 at each temperature.
At fixed $u_r$ in this range, $K$ varies by at most 30\% across the four temperatures.
Its dominant dependence in this window is therefore on $u_r$.
At larger $u_r$, the finite range reduces $K$ at the highest temperature.

The two-mode factor of Eq.~\eqref{eq:K2} stays close to unity near $u_r=4$.
Its minimum $K^{(2)}=1$ occurs at the mean collision energy $\bar u=4$, where the direct coupling $L_{35}$ between the heat current and the fifth moment vanishes.
It thus accounts for the weak memory up to $u_r\approx4$, but not for the magnitude of the growth beyond.
At $u_r=5.75$ and $\vartheta=0.01$, it gives $K^{(2)}=1.40$, whereas the three-mode truncation gives $K^{(3)}=2.91$ and the full 96-mode value is $K=6.04$.
The slow relaxation therefore involves several odd moments.

For the 96-mode calculation, let $v_\alpha$ be orthonormal eigenmodes of the projected collision operator with relaxation rates $\lambda_\alpha$.
The finite-basis form of Eq.~\eqref{eq:Ca} is
\begin{equation}
 \begin{aligned}
 C_3(t)&=\sum_\alpha w_\alpha\e^{-\lambda_\alpha t},\\
 w_\alpha&=|(\psi_3,v_\alpha)_0|^2,\qquad
 \sum_\alpha w_\alpha=1.
 \end{aligned}
 \label{eq:discrete_spectrum}
\end{equation}
Here $w_\alpha$ is the heat-current spectral weight of mode $v_\alpha$, and the sum runs over all 96 modes.
Equations~\eqref{eq:initial} and \eqref{eq:transport} then give
\begin{equation}
 \gamma_Q=\sum_\alpha w_\alpha\lambda_\alpha,\qquad
 \gamma_\tr^{-1}=\sum_\alpha\frac{w_\alpha}{\lambda_\alpha}.
 \label{eq:discrete_rates}
\end{equation}
The inverse-rate weighting makes the integrated time sensitive to slow modes with appreciable heat-current weight.
Figure~\ref{fig:memory}(b) shows how this weight is redistributed.
For $1\le u_r\le3.5$, more than two thirds of the weight lies within a factor of 2 of $\gamma_Q$.
For $u_r\gtrsim4$, one branch carries an increasing part of the weight to slow rates, while the remaining weight moves above $\gamma_Q$.
At $u_r=5.75$, the slowest mode has $\lambda_s=0.065\gamma_Q$ and carries 33\% of the heat-current weight.
At $u_r=3.75$, the slowest mode has $\lambda_s=0.20\gamma_Q$ and carries only 8\%.
The fast branch maintains the mean $\gamma_Q$, whereas the slow branch dominates the integrated time $\gamma_\tr^{-1}$.

The fixed-energy collision geometry identifies the origin of the slow branch.
We compare its eigenvector with those of the 96-mode shell matrix $\mathbf G(u_r)$ at the resonance energy.
At $\vartheta=0.01$, their squared overlap lies between 0.91 and 0.99 for $3\le u_r\le10$, and at $\vartheta=0.003$ it exceeds 0.99 for $u_r=5.75$ and 8.
At $u_r=5.75$ and $\vartheta=0.01$, this shell mode has the rate $0.061\gamma_Q$ and the heat-current weight 0.35, close to the full values $0.065\gamma_Q$ and 0.33.
Resonant collisions within 16 half-widths of the resonance supply 94\% of the decay rate of the slow mode, nearly the same fraction as for $\gamma_Q$.
The decay of this mode is dominated by resonant collisions.

Equation~\eqref{eq:shell} shows how such a mode arises.
For the third and fifth moments, the changes are proportional to $\Delta\Sigma_3$, with different center-dependent prefactors.
The seventh moment adds a term in $\Delta[(\Sigma_3)^2]$ (Appendix~\ref{app:gram}).
A combination of $\psi_3$, $\psi_5$, and $\psi_7$ can then reduce the collision-induced change.
At $u_r=5.75$, the slow mode consists of 33\% $\psi_3$, 38\% $\psi_5$, and 16\% $\psi_7$.
Its small decay rate reflects cancellations among these moment changes.
In momentum space, the slow mode is concentrated at $|y|\lesssim2.5$.
A slow particle reaches the resonance energy only with partners from the thermal tail.
The single-particle loss frequency is the rate at which collisions scatter a particle out of its momentum state.
At $y=0$, this frequency is $0.12\gamma_Q$ for $u_r=5.75$, compared with $0.47\gamma_Q$ for $u_r=3.75$.

\begin{figure*}[t!]
 \includegraphics[width=\textwidth]{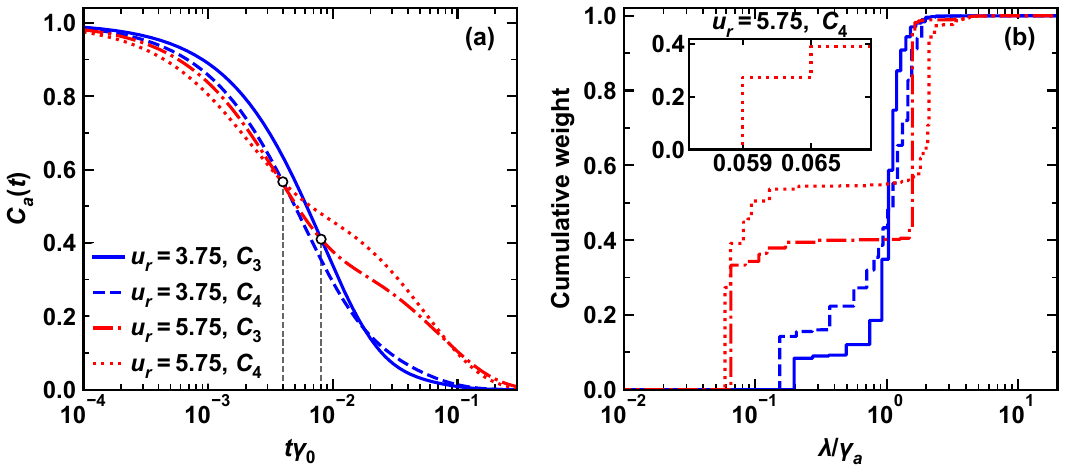}
 \caption{Finite-time relaxation at $\vartheta=0.01$.
 Blue and red curves correspond to $u_r=3.75$ and 5.75, respectively.
 (a) Heat-current and fourth-moment correlations on the common clock $t\gamma_0$.
 Open circles mark the red--blue intersections for each source.
 Gray dashed vertical guides project the intersections onto the time axis.
 (b) Cumulative spectral weights of the same sources, with each spectrum scaled by its own initial rate.
 Each cumulative weight sums the source weight at rates not exceeding $\lambda$.
 The inset resolves the first two steps of the red $C_4$ curve.
 The line styles are the same in both panels.}
 \label{fig:relaxation}
\end{figure*}

\subsection{Finite-time relaxation}

Figure~\ref{fig:relaxation} shows the time-domain consequence for the two representative states.
Both the heat-current and fourth-moment correlations initially decrease faster at $u_r=5.75$ (red curves) than at $u_r=3.75$ (blue curves).
Their initial decay rates are larger by a factor of 1.61.
At later times, the red curves lie above the blue curves over a substantial time window.
In the 96-mode calculation, the $C_3(t)$ curves at $u_r=3.75$ and 5.75 cross near $t\gamma_0=8.05\times10^{-3}$.
The corresponding $C_4(t)$ curves cross near $t\gamma_0=3.97\times10^{-3}$.
Figure~\ref{fig:relaxation}(b) shows the corresponding spectral weights.
At $u_r=5.75$, the fourth-moment source has its two slowest modes at $\lambda=0.059\gamma_4$ and $0.065\gamma_4$, with a combined weight of 39\%.
The inset resolves these two steps in the red dotted curve.
At $u_r=3.75$, neither source has weight below $0.15\gamma_a$.

For a quantitative comparison, the decay times $t_{e,a}$ and $t_{10,a}$ mark the decrease of $C_a$ to $1/\e$ and 0.1,
\begin{equation}
 C_a(t_{e,a})=\e^{-1},\qquad C_a(t_{10,a})=0.1.
 \label{eq:times}
\end{equation}
Table~\ref{tab:times} lists both criteria.
Compared with $u_r=3.75$, the fourth-moment $1/\e$ time at $u_r=5.75$ is larger by a factor of 2.87.
Its time to reach 0.1 is larger by a factor of 3.48.
The heat-current times also increase.
The state with the larger initial rate therefore recovers more slowly by both criteria and for both probes.

\begin{table}[t]
 \caption{Initial rates and finite-time decay criteria at $\vartheta=0.01$.
 The source index is $a=3$ for heat and $a=4$ for the fourth moment.
 All times use the common reference $\gamma_0$.
 The correlations are calculated with 96 modes.}
 \label{tab:times}
 \setlength{\tabcolsep}{3pt}
 \begin{ruledtabular}
 \begin{tabular}{ccrrr}
 $u_r$ & $a$ & $\gamma_a/\gamma_0$ & $t_{e,a}\gamma_0$ & $t_{10,a}\gamma_0$\\
 \hline
 3.75 & 3 &117.8&0.00912&0.0247\\
 3.75 & 4 &157.0&0.00768&0.0287\\
 5.75 & 3 &189.2&0.0107&0.103\\
 5.75 & 4 &252.3&0.0220&0.0996\\
 \end{tabular}
 \end{ruledtabular}
\end{table}

The two probes also have different orderings within one state.
At $u_r=5.75$, the fourth moment takes about twice as long as the heat current to reach $1/\e$, although its initial rate is larger by the exact factor $4/3$.
At the 0.1 criterion, their ordering is reversed.
Through Eq.~\eqref{eq:entropy}, the fourth-moment correlation also gives the quadratic entropy recovery of the initially specified distortion.

The even sector is integrated independently with the 96 basis functions $\psi_4,\psi_6,\ldots,\psi_{194}$.
For the energy integral over $u=E/T$, we use 48 Gauss--Legendre nodes per subinterval with an upper limit $u_{\max}=550$.
We repeat the calculation with 72 nodes per subinterval and extend the upper limit to $u_{\max}=660$.
This comparison tests stability under both finer quadrature and a larger energy cutoff.
The largest scaled matrix change is below $2.4\times10^{-12}$, and the largest 80-to-96-mode relative change in the decay times is below $4.7\times10^{-7}$.

\subsection{Finite-frequency heat response}

\begin{figure*}[t!]
 \includegraphics[width=\textwidth]{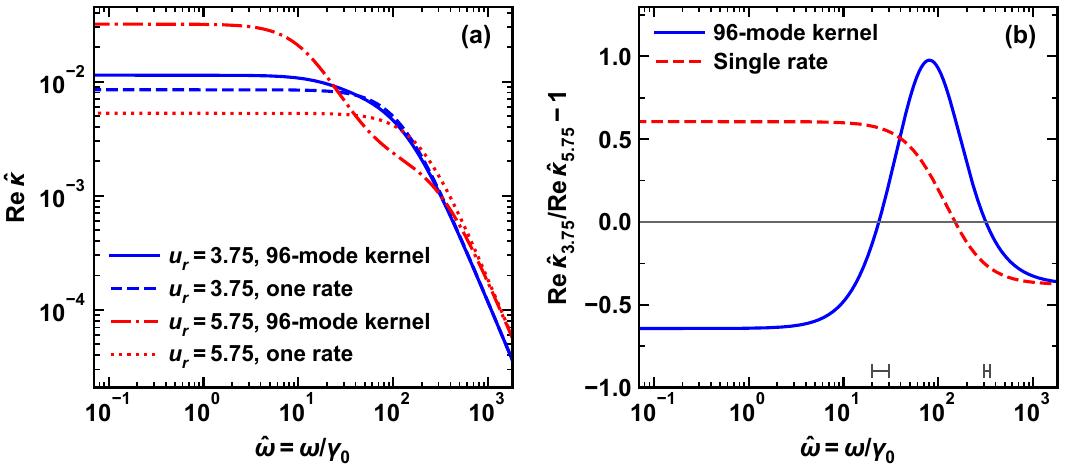}
 \caption{Finite-frequency heat response at $\vartheta=0.01$.
 (a) The dissipative response at the two representative detunings.
 Solid and dash-dotted curves use the collision kernel projected onto 96 odd Hermite modes.
 Dashed and dotted curves use a single rate fixed by the corresponding initial slope.
 (b) The response ratio minus one.
 Horizontal brackets near the lower axis mark the certified intervals $20<\hat\omega<30$ and $300<\hat\omega<350$, each containing at least one crossing.
 The single-rate comparison has only one crossing.}
 \label{fig:frequency}
\end{figure*}

The dimensionless thermal conductivity $\hat\kappa(\hat\omega)$ is obtained from the scaled collision operator $\hat{\mathcal L}=\mathcal L/\gamma_0$ as
\begin{equation}
 \hat\kappa(\hat\omega)
 =(\psi_3,(\hat{\mathcal L}-\ii\hat\omega)^{-1}\psi_3)_0,\qquad
 \hat{\mathcal L}=\mathcal L/\gamma_0.
 \label{eq:dimensionless}
\end{equation}
The real part $\Ree\hat\kappa(\hat\omega)$ is the dissipative heat response shown in Fig.~\ref{fig:frequency}(a).
For each state, this response decreases monotonically with increasing frequency.
The transport rate $\gamma_\tr$ defined in Eq.~\eqref{eq:transport} fixes the zero-frequency value, $\Ree\hat\kappa(0)=\gamma_0/\gamma_\tr$.
The initial heat-current decay rate $\gamma_Q$ defined in Eq.~\eqref{eq:initial} fixes the high-frequency tail, $\Ree\hat\kappa\simeq(\gamma_Q/\gamma_0)\hat\omega^{-2}$.
At $u_r=5.75$, $\gamma_Q$ is larger and $\gamma_\tr$ is smaller than at $u_r=3.75$.
The response at $u_r=5.75$ is therefore larger in both the low- and high-frequency regimes.

Figure~\ref{fig:frequency}(a) shows the dissipative responses.
At low frequencies, the response at $u_r=5.75$ is 2.8 times larger.
At intermediate frequencies, the response at $u_r=3.75$ becomes larger, because the weight of the $u_r=5.75$ state lies either well below or well above this range.
At high frequencies, the ordering returns.
The 96-mode estimates of the two crossings are $\hat\omega=23.6$ and 316.
Near $\hat\omega=100$, the dissipative heat response at $u_r=3.75$ is nearly twice that at $u_r=5.75$.

A single-rate description cannot produce this behavior.
For a single-rate model with positive relaxation rate $\gamma$, the dissipative response $\Ree\hat\kappa^{(1)}$ is
\begin{equation}
 \Ree\hat\kappa^{(1)}(\hat\omega)
 =\frac{\hat\gamma}{\hat\gamma^2+\hat\omega^2},\qquad
 \hat\gamma=\gamma/\gamma_0,
\end{equation}
and the difference between two states with rates $\hat\gamma$ and $\hat\gamma'$ obeys
\begin{equation}
 \Ree\hat\kappa^{(1)}-\Ree\hat\kappa'^{(1)}
 =\frac{(\hat\gamma-\hat\gamma')(\hat\omega^2-\hat\gamma\hat\gamma')}
 {(\hat\gamma^2+\hat\omega^2)(\hat\gamma'^2+\hat\omega^2)}.
 \label{eq:single}
\end{equation}
For distinct rates, the two limits have opposite orderings, and there is exactly one positive crossing at $\hat\omega=(\hat\gamma\hat\gamma')^{1/2}$.
Different positive prefactors leave the numerator linear in $\hat\omega^2$ and still allow at most one crossing.
With the initial slopes, the single-rate crossing lies at $\hat\omega=149$, as shown in Fig.~\ref{fig:frequency}(b).
Truncating the kernel to two odd modes also gives only one crossing, whereas three odd modes already give two.

\begin{table}[t]
 \caption{Full-operator enclosures of the response difference $\Delta\hat\kappa$.
 Endpoints are rounded outward.
 Opposite signs at 20 and 30 and at 300 and 350 establish two disjoint crossing intervals.}
 \label{tab:certificates}
 \begin{ruledtabular}
 \begin{tabular}{cl}
 $\hat\omega$ & Enclosure of $\Delta\hat\kappa(\hat\omega)$\\
 \hline
 0.1 & $[-2.17,-1.94]\times10^{-2}$\\
 1   & $[-2.05,-2.02]\times10^{-2}$\\
 10  & $[-9.94,-9.89]\times10^{-3}$\\
 20  & $[-1.47,-1.44]\times10^{-3}$\\
 30  & $[1.57,1.59]\times10^{-3}$\\
 300 & $[3.42,3.43]\times10^{-5}$\\
 350 & $[-5.09,-5.08]\times10^{-5}$\\
 \end{tabular}
 \end{ruledtabular}
\end{table}

The orderings also hold for the full operator.
We denote the difference between the dissipative responses at the two representative detunings by
\begin{equation}
 \Delta\hat\kappa(\hat\omega)=
 \Ree\hat\kappa_{3.75}(\hat\omega)
 -\Ree\hat\kappa_{5.75}(\hat\omega).
\end{equation}
Table~\ref{tab:certificates} lists outward-rounded enclosures.
Among the sampled frequencies, the largest full width of the certified response-ratio intervals is below $3.23\times10^{-2}$ at $\hat\omega=0.1$.
At $\hat\omega=1$, the certified ratio $\Ree\hat\kappa_{3.75}/\Ree\hat\kappa_{5.75}$ lies between 0.356 and 0.361.
At $\hat\omega=1$, the response at $u_r=5.75$ is larger than at $u_r=3.75$ by a factor between 2.77 and 2.81.
At $\hat\omega=0.1$, this factor lies between 2.67 and 2.94.
The signs at 20 and 30 and at 300 and 350 imply at least one crossing in each interval, since the response is continuous at nonzero frequency.
At $\hat\omega=100$, the certified ratio lies between 1.926 and 1.927.

Appendix~\ref{app:bounds} describes the construction of these enclosures.
It combines polynomial trial functions with a residual bound.
Every uncomputed Hermite row is included through an analytical tail estimate, and the energy integrals are enclosed with ball arithmetic.
The enclosures therefore control the infinite collision operator within the stated kinetic model.

\subsection{Range of validity}
\label{sec:validity}

The vacuum amplitude is treated nonperturbatively in $g_3$, and the gas dynamics is retained at leading order in fugacity.
Pauli blocking and the quantum-statistical corrections to $f_{k,0}$ are of relative order $z=0.03$, and medium corrections to the amplitude enter at higher order in density.
For all states, the interparticle distance exceeds the range $1/\Lambda$ by a factor of at least 800.
Two-body collisions alone do not relax the momentum distribution, but an odd-wave two-body interaction would add rescattering to the three-body amplitude.
The present model omits this background and applies when the resonant three-body channel dominates the amplitude.
The narrow-resonance thermal weight $u_r^6\e^{-u_r}$ follows from the threshold law and the Boltzmann factor.
With the full finite-range amplitude, the grid maximum of $\gamma_Q$ shifts from $u_r=6$ to 5.5 across the four temperatures.
The transport maximum shifts from $u_r=4$ to 3.75, one grid step.

The Wigner time delay $t_W$ defined in Eq.~\eqref{eq:delay} supplies a consistency check.
For $u_r=3.75$ and 5.75 at $\vartheta=0.01$, the products are, respectively,
\begin{equation}
 \gamma_Qt_W=4.39\times10^{-4},\quad
                          2.08\times10^{-4}.
\end{equation}
At $\hat\omega=350$, the corresponding frequency products are
\begin{equation}
 \omega t_W=1.30\times10^{-3},\quad
                         3.85\times10^{-4}.
\end{equation}
The collision delay is thus short compared with the relaxation and probing times considered here.
Resolving dynamics on the scale of the collision delay would require an explicit treatment of collision memory.
For all states studied, $\eta\le0$, so the separable channel has no trimer below the three-body threshold and the kinetics contains no cluster population.

The present calculation describes homogeneous kinetic relaxation and the associated heat response.
In one dimension, spatial fluctuations of particle, momentum, and energy densities can generate long-time tails in heat-current correlations~\cite{Narayan2002,Lepri2003,Dhar2008}.
Nonlinear fluctuating hydrodynamics describes their coupled dynamics~\cite{Mendl2013,Spohn2014}.
These long-time tails can modify the zero-frequency thermal conductivity.
A quantitative treatment of their contribution lies beyond the scope of the present work.

\section{Summary and perspectives}
\label{sec:summary}

We have studied how a three-body resonance controls kinetic relaxation and heat transport in a dilute one-dimensional gas of identical fermions.
We derived the vacuum amplitude for a finite-range interaction with the leading antisymmetric momentum structure.
Its full energy dependence enters the linearized Boltzmann collision operator used to calculate the relaxation spectra and heat response.

The initial heat-current decay rate is largest for a resonance near $u_r=6$, whereas integrated heat relaxation is fastest near $u_r=4$.
The first scale follows from the threshold law of the antisymmetric vertex.
The second reflects how resonant collisions redistribute the heat current among odd moments.
For $1\le u_r\le4$, the memory factor remains below 2.
At higher $u_r$, a slow mode carries a substantial fraction of the heat-current weight.
This mode closely follows an eigenmode of the collision matrix evaluated at the resonance energy.
At the initial-rate maximum, the memory factor lies between 4.6 and 7.7 for the temperatures studied.

A larger initial decay rate can therefore coexist with slower integrated relaxation and a larger low-frequency heat response.
At $\vartheta=0.01$, moving the resonance from $u_r=3.75$ to 5.75 raises the initial rate by 61\%.
The heat response at $\hat\omega=1$ increases by a factor between 2.77 and 2.81.
The two dissipative responses reverse their ordering at least twice as the frequency increases.
Residual bounds on the full collision operator establish both reversals.
A single-rate description gives at most one crossing.

A fourth-moment distortion provides an independent probe of the same relaxation dynamics.
Its initial rate is exactly $4/3$ of the heat-current rate for any Galilean-invariant three-body collision kernel.
Like the heat current, it can decay faster initially yet recover more slowly at later times.
The associated kurtosis can be obtained from momentum distributions.
The fourth-moment correlation also determines the quadratic entropy deficit of the specified initial distortion.

For future perspectives, the kinetic framework can be extended to include odd-wave two-body interactions, Pauli blocking, and medium corrections.
These extensions would reveal how two-body rescattering and quantum statistics modify the resonance-controlled relaxation.
A coupling to nonlinear fluctuating hydrodynamics would connect the kinetic response to heat transport at long wavelengths.

\begin{acknowledgments}
Y.G. is supported by RIKEN Special Postdoctoral Researchers Program.
\end{acknowledgments}

\appendix
\section{Amplitude normalization and low-energy matching}
\label{app:norm}

The inverse Jacobi transformation has the measure
\begin{equation}
 \dd k_1\dd k_2\dd k_3=\frac1{\sqrt3}\dd P\dd p\dd q.
\end{equation}
With our relative-state normalization, the product-state amplitude therefore contains a factor $\sqrt3$.
Antisymmetrizing the three-particle states gives a further factor $3!=6$, because $F$ is antisymmetric on both legs.
This gives the physical amplitude in Eq.~\eqref{eq:M}.
The collision prefactor $1/12$ is multiplied by $1/3$ when averaging over incoming legs, and symmetrizing incoming and outgoing triples gives another factor $1/2$.
Their product is the factor $1/72$ in Eq.~\eqref{eq:quadratic}.

For the hypervolume matching, we write the relative coordinate conjugate to $\bm\rho$ as $(R\cos\chi,R\sin\chi)$.
Here $R$ and $\chi$ are the relative-coordinate hyperradius and hyperangle, and $\Psi$ denotes the zero-energy scattering wave function.
The incident zero-energy wave function is $R^3\cos3\chi/\sqrt2$, and the free Green function is $m\ln R/\pi$ up to a constant.
Applying the derivative polynomial gives
\begin{equation}
 \Psi\simeq\frac{R^3\cos3\chi}{\sqrt2}\left[1-\frac{96m\tau(0)}{\pi R^6}\right].
\end{equation}
The Gaussian factor affects only shorter distances in this matching.
In the convention of Ref.~\cite{Wang2023}, the hyperradius is $B=\sqrt{3/2}\,R$ and
\begin{equation}
 \Psi\simeq\frac{R^3\cos3\chi}{\sqrt2}\left[1-\frac{3\sqrt3D_F}{2\pi B^6}\right].
\end{equation}
Matching the two expressions gives Eq.~\eqref{eq:hypervolume}.
It also gives the low-energy scattering amplitude and the three-particle energy shift $\Delta E_3$ in a periodic box,
\begin{equation}
 M_{456;123}\simeq\frac{D_F}{12m}X_{\rm out}X_{\rm in},\qquad
 \Delta E_3=\frac{D_FX_{\rm in}^2}{12m\ell^2},
\end{equation}
where $X=(k_1-k_2)(k_2-k_3)(k_3-k_1)$ is the antisymmetric momentum polynomial and $\ell$ is the length of a periodic box.
The energy shift agrees with Ref.~\cite{Wang2023}.

\section{Positive construction of collision matrices}
\label{app:gram}

We use the relative-momentum hyperangle $\alpha$, with $p=\rho\cos\alpha$ and $q=\rho\sin\alpha$.
The vertex then has the angular factor $\cos3\alpha$.
At fixed $u$, the particle momenta relative to the center are written using fixed angular offsets $\alpha_i$ as
\begin{align}
 \xi_i&=\sqrt{4u/3}\cos(\alpha-\alpha_i),\\
 (\alpha_1,\alpha_2,\alpha_3)&=(\pi/6,5\pi/6,3\pi/2),
\end{align}
and the centers have the density $\sqrt{3/(2\pi)}\e^{-3s^2/2}$.
Let $\Delta_j$ be the incoming minus outgoing value of $\sum_i\He_j(y_i)$.
Then
\begin{align}
 G_{jj'}(u)={}&\frac{9}{\pi^2u^3\sqrt{j!j'!}}
 \int\dd s\sqrt{\frac3{2\pi}}\e^{-3s^2/2}
 \nonumber\\
 &\times\int_0^{2\pi}\dd\alpha\,\dd\alpha'\,
 \cos^23\alpha\cos^23\alpha'\,\Delta_j\Delta_{j'},
 \label{eq:Ggeneral}
\end{align}
which is normalized to $G_{33}=1$.
Direct integration gives Eqs.~\eqref{eq:oddblock} and \eqref{eq:evenblock}.
The three-mode truncation retains $\psi_3$, $\psi_5$, and $\psi_7$, the first three odd basis functions outside the conserved space.
The additional matrix elements used in Sec.~\ref{sec:slow} are
\begin{subequations}
\label{eq:Gseven}
\begin{align}
 G_{37}&=\frac{\sqrt{210}}{180}(u^2-10u+20),\\
 G_{57}&=\frac{\sqrt{42}}{216}(u-4)(u^2-10u+40),\\
 G_{77}&=\frac{7}{9720}(9u^4-179u^3+1710u^2\nonumber\\
 &\qquad-7200u+11400).
\end{align}
\end{subequations}
For the seventh moment, the single-collision change is
\begin{align}
 \Delta\sum_i\He_7(y_i)={}&\Big[35s^4+(35u-210)s^2\nonumber\\
 &+\tfrac73u^2-35u+105\Big]\Delta\Sigma_3
 +\tfrac73s\,\Delta[(\Sigma_3)^2],
\end{align}
which includes a contribution from the squared cubic sum $(\Sigma_3)^2$.

For high orders, we introduce the relative Hermite polynomials $p_h$ and their three-particle sums $\bar p_h$,
\begin{align}
 p_h(\alpha)&=\frac{(2/3)^{h/2}}{\sqrt{h!}}
 \He_h(\sqrt{2u}\cos\alpha),\\
 \bar p_h(\alpha)&=\sum_i p_h(\alpha-\alpha_i).
\end{align}
Hermite addition gives
\begin{equation}
 \sum_i\psi_j(y_i)=
 \sum_{l=0}^j\sqrt{\binom jl}\,3^{-l/2}
 \psi_l(\sqrt3s)\,\bar p_{j-l}(\alpha).
\end{equation}
Here $\binom jl=j!/[l!(j-l)!]$ is the binomial coefficient.
The center integral follows from orthogonality, and only $j-l\ge3$ contributes.
The Fourier coefficients of $p_h$ follow from the recurrence
\begin{equation}
 p_{h+1}=\frac{\sqrt{4u/3}\cos\alpha}{\sqrt{h+1}}p_h
       -\frac23\sqrt{\frac h{h+1}}p_{h-1},\qquad p_0=1.
\end{equation}
Writing the Fourier coefficients as $p_{hn}$ in $p_h=\sum_np_{hn}\e^{\ii n\alpha}$, we introduce auxiliary angular coefficients $c_{hn}=3(-\ii)^{n/3}p_{hn}$ for nonzero $n$ divisible by 3.
We set $c_{hn}=0$ for other nonzero $n$.
The angular mean is removed by $c_{h0}=-(c_{h6}+c_{h,-6})/2$.
The coefficients $d_{hn}=(c_{h,n-3}+c_{h,n+3})/2$ include the vertex angular factor.
Parseval's identity then gives the positive Gram representation
\begin{align}
 G_{jj'}(u)={}&\frac{36}{u^3}
 \sum_{l=0}^{\min(j,j')-3}
 \sqrt{\binom jl\binom{j'}l}\,3^{-l}
 \nonumber\\
 &\times\Ree\sum_n d_{j-l,n}d_{j'-l,n}^*.
\end{align}
This construction avoids subtracting separately integrated large moments.
Independent angular and Gaussian quadrature checks the low-order elements, and exact rational polynomial generation supplies a further high-order check.

The remaining energy integral uses piecewise Gaussian quadrature.
Near the crossing, a tangent substitution with the scale $\Gamma_{\rm lin}/2T$ maps the resonance onto a smooth integrand.
The full amplitude is evaluated at every node.
For the odd sector, we successively use 32, 48, and 64 Gauss--Legendre nodes per energy subinterval.
A final convergence check uses 72 nodes per subinterval and increases the upper limit of $u=E/T$ from 436 to 545.
This checks convergence with respect to both quadrature order and the energy cutoff.
An independent adaptive scalar integral checks $Z$.
Basis convergence is tested through 16, 32, 48, 64, 80, and 96 modes.

\section{Bounds on the full response}
\label{app:bounds}

This construction gives enclosures for the heat response of the full operator $\hat{\mathcal L}=\mathcal L/\gamma_0$ at $\hat\omega>0$.
An enclosure is an interval guaranteed to contain the exact real response of the kinetic model.
The odd complex polynomial trial function $\phi$ approximates the heat-response solution $(\hat{\mathcal L}-\ii\hat\omega)^{-1}\psi_3$.
We expand it as $\phi(y)=\sum_{j=3,5,\ldots,J}\phi_j\psi_j(y)$, with $\phi_j=(\psi_j,\phi)_0$ and maximum Hermite order $J$.
We define its kinetic-equation residual $\varepsilon_\phi$ and response estimate $\hat\kappa_\phi$ by
\begin{align}
 \varepsilon_\phi&=\psi_3-(\hat{\mathcal L}-\ii\hat\omega)\phi,\label{eq:kineticresidual}\\
 \hat\kappa_\phi&=2(\psi_3,\phi)_0-(\phi^*,(\hat{\mathcal L}-\ii\hat\omega)\phi)_0.
\end{align}
The superscript $*$ denotes complex conjugation.
The inner product already conjugates its first argument, so $(\phi^*,\cdot)_0$ contains $\phi$ without conjugation in the integrand.
Since $\hat{\mathcal L}$ is real and symmetric,
\begin{equation}
 \hat\kappa-\hat\kappa_\phi
 =(\varepsilon_\phi^*,(\hat{\mathcal L}-\ii\hat\omega)^{-1}\varepsilon_\phi)_0.
\end{equation}
Because $\hat{\mathcal L}$ is nonnegative,
\begin{equation}
 |\hat\kappa-\hat\kappa_\phi|\le\frac{\|\varepsilon_\phi\|^2}{\hat\omega},
 \label{eq:residual}
\end{equation}
where $\|\varepsilon_\phi\|^2=(\varepsilon_\phi,\varepsilon_\phi)_0$ is the squared norm of the residual.
Even for a finite polynomial trial, $\hat{\mathcal L}\phi$ generally contains higher Hermite orders.
Every omitted component must be included in this residual.

\subsection{Gaussian action and Hermite tail}

Using $\sin^2\delta_3=b^2/(a^2+b^2)$, the matrix elements $\hat L_{jj'}$ of the dimensionless collision operator have the energy representation
\begin{align}
 \hat L_{jj'}&=\frac{\vartheta^2}{7!}\int_0^\infty\dd u\,w(u)G_{jj'}(u),\\
 \frac{\vartheta^2w(u)}{7!}&=\frac{u^3\e^{-u}\sin^2\delta_3}
                         {5040\pi^2\vartheta^4},
 \label{eq:ru}
\end{align}
where $\hat L_{jj'}=L_{jj'}/\gamma_0$, and $\delta_3$ is evaluated at $E=uT$.
The quadratic form and unitarity, $0\le\sin^2\delta_3\le1$, imply
\begin{equation}
 0\le\hat{\mathcal L}\le\lambda_{\max},\qquad
 \lambda_{\max}=\frac9{70\pi^2\vartheta^4}.
\end{equation}
Here $\lambda_{\max}$ is a uniform upper bound on the dimensionless relaxation rates.
With $\sin^2\delta_3\le1$ and $\sin^23\alpha\le1$ in Eq.~\eqref{eq:action} below, the loss frequency of a single particle is at most $\lambda_{\max}/6$ at every momentum.
The inequality $|\Delta\phi|^2\le6\sum_{i=1}^6|\phi(k_i)|^2$ then gives the bound.
In dimensional form, $\gamma_0\lambda_{\max}=144\sqrt3 n^2/m$.

For the polynomial action, we shift the hyperangle by $\pi/6$ and cyclically relabel the particles, so the angular weight becomes $\sin^23\alpha$.
We use the rotated relative momenta $\xi_i=\sqrt{4u/3}\cos(\alpha-2\pi i/3)$ for $i=1,2,3$.
The collision-induced change averaged over the outgoing hyperangle is
\begin{align}
 \bar\Delta_\phi(s,u,\alpha)={}&\sum_i\phi(s+\xi_i(\alpha))
 \nonumber\\
 &-\frac1\pi\int_0^{2\pi}\dd\alpha'\,
      \sin^23\alpha'\sum_i\phi(s+\xi_i(\alpha')).
\end{align}
The bar denotes the outgoing-angle average.
The prime labels the outgoing hyperangle $\alpha'$.
The full action is then
\begin{align}
 (\hat{\mathcal L}\phi)(y)={}&\frac{3\sqrt3}{280\pi^3\vartheta^4}
 \int_0^\infty\dd u\,\e^{-u}\sin^2\delta_3
 \nonumber\\
 &\times\int_0^{2\pi}\dd\alpha\,\sin^23\alpha\,
 \e^{-y^2+3\xi_3y-3\xi_3^2/2}
 \nonumber\\
 &\times\bar\Delta_\phi(y-\xi_3,u,\alpha).
 \label{eq:action}
\end{align}
This follows by conditioning on the incoming momentum $y=s+\xi_3$, and the two partner occupations give the Gaussian factor.

With the normalized equilibrium momentum weight $w_0(y)=\e^{-y^2/2}/\sqrt{2\pi}$, the Hermite generating transform of a function $f(y)$ is
\begin{equation}
 \tilde f(\zeta)=\int\dd y\,w_0(y)f(y)\e^{\zeta y-\zeta^2/2}
 =\sum_j\frac{(\psi_j,f)_0\,\zeta^j}{\sqrt{j!}}.
\end{equation}
The tilde denotes this generating transform, and $\zeta$ is its auxiliary complex variable.
The transform of $\hat{\mathcal L}\phi$ is entire, meaning analytic throughout the complex $\zeta$ plane.
The basic Gaussian identity is
\begin{align}
 &\int\dd y\,w_0(y)\e^{\zeta y-\zeta^2/2}
 \nonumber\\
 &\quad\times\e^{-y^2+3\xi_3y-3\xi_3^2/2}\psi_l(\sqrt3(y-\xi_3))
 \nonumber\\
 &\hspace{15mm}=
 \frac{\e^{\xi_3\zeta-\zeta^2/3}}{\sqrt3}
 \frac{\zeta^l}{3^{l/2}\sqrt{l!}}.
\end{align}
These ingredients give the growth bound
\begin{equation}
 |\widetilde{\hat{\mathcal L}\phi}(\zeta)|
 \le\frac{3\lambda_{\max}}{2}\e^{(\Imm\zeta)^2/3}U_\phi(|\zeta|),
 \label{eq:growth}
\end{equation}
where $\Imm\zeta$ denotes the imaginary part of $\zeta$.
The auxiliary polynomials $U_\phi$ and $V_h$ have nonnegative coefficients and are defined by
\begin{align}
 U_\phi(R)={}&\sum_{j\le J}|\phi_j|
 \sum_{l=0}^{j-3}
 \frac{\sqrt{\binom jl}}{3^l\sqrt{l!}}
 \left(\frac23\right)^{(j-l)/2}R^l
 V_{j-l}(R),\\
 V_h(R)={}&\sum_{l'=0}^h
 \sqrt{\frac{\binom h{l'}}{l'!}}\left(\sqrt{\frac23}R\right)^{l'} .
\end{align}
Here $h$ is a nonnegative integer, and $l,l'$ are integer summation indices.
The prime on $l'$ labels a distinct index.
The degree of $U_\phi$ is at most $J$.
The argument $R$ denotes a nonnegative radius in the complex $\zeta$ plane.

The bound follows in a few steps.
Hermite addition expands each $\phi(s+\xi_i)$ into products of $\psi_l(\sqrt3s)$ and $\He_h(Z_i)/\sqrt{h!}$, with $Z_i=\sqrt{3/2}\,\xi_i$ and $h=j-l$, and the Gaussian identity turns $\psi_l(\sqrt3s)$ into $\zeta^l$.
The terms with $h\le2$ are conserved and cancel in $\bar\Delta_\phi$.
The measure $\e^{-u}\dd u\,\dd\alpha$ is a two-dimensional Gaussian of total weight $2\pi$, under which each $Z_i$ is a standard normal variable.
We bound $\sin^2\delta_3$ and $\sin^23\alpha$ by unity.
The factor $\exp(\xi_3\Ree\zeta)$ in $|\e^{\xi_3\zeta-\zeta^2/3}|$ shifts this Gaussian, leaves the factor $\e^{(\Imm\zeta)^2/3}$, and displaces each $Z_i$ by at most $\sqrt{2/3}\,|\zeta|$.
The outgoing average is a uniform rotation of the shifted Gaussian and obeys the same displacement bound.
For a standard normal variable $Z$, we write $\langle g(Z)\rangle=\int\dd Z\,w_0(Z)g(Z)$ for the average of a function $g$.
The identity $\He_h(z+c)=\sum_{q=0}^h\binom hq c^{h-q}\He_q(z)$ and the Cauchy--Schwarz bound $\langle|\He_q(Z)|\rangle\le\sqrt{q!}$ give $\langle|\He_h(Z+c)|\rangle\le\sqrt{h!}\,V_h(|\zeta|)$ for $|c|\le\sqrt{2/3}\,|\zeta|$.
The incoming triple supplies three such terms, and the outgoing average at most six more, because its angular density $\sin^23\alpha'/\pi$ is bounded by twice the uniform density.
The resulting factor $9\times2\pi/\sqrt3$ times the prefactor of Eq.~\eqref{eq:action} equals $3\lambda_{\max}/2$.

For the Hermite coefficients $c_j=(\psi_j,\hat{\mathcal L}\phi)_0$ of the operator action, Cauchy's estimate on a circle of radius $\sqrt{3j/2}$ gives
\begin{equation}
 |c_j|^2\le
 j!\left(\frac{3\lambda_{\max}}2\right)^2U_\phi\!\left(\sqrt{3j/2}\right)^2
 \e^j\left(\frac2{3j}\right)^j.
\end{equation}
For an odd trial, let $j_0>J$ be the first odd Hermite order omitted from the explicit residual calculation.
The resulting geometric tail bound $\Theta_{j_0}(\phi)$ obeys
\begin{align}
 \sum_{\substack{j\ge j_0\\j\ {\rm odd}}}|c_j|^2&\le\Theta_{j_0}(\phi),\\
 \Theta_{j_0}(\phi)&=
 \frac{\e j_0 (3\lambda_{\max}/2)^2U_\phi(\sqrt{3j_0/2})^2(2/3)^{j_0}}
 {1-(4/9)(1+2/j_0)^{J+1}},
 \label{eq:tail}
\end{align}
where the denominator is required to be positive.
The estimate uses $j!\le\e j(j/\e)^j$ and the nonnegative polynomial coefficients.
It bounds all uncomputed Hermite orders.

\subsection{Interval integrals and numerical enclosure}

The scaled energy moments $Z_r$ of nonnegative integer order $r$ are
\begin{equation}
 Z_r=\frac{\vartheta^2}{7!\,(r+3)!}\int_0^\infty\dd u\,w(u)u^r.
\end{equation}
We denote the lower and upper integration endpoints by $u_{\min}$ and $u_{\max}$.
The superscripts in $Z_r^{(0,u_{\min})}$ and $Z_r^{(u_{\max},\infty)}$ specify the energy intervals contributing to $Z_r$.
These low- and high-energy contributions are nonnegative and satisfy
\begin{align}
 Z_r^{(0,u_{\min})}
 &\le\frac{u_{\min}^{r+4}}
 {5040\pi^2\vartheta^4(r+4)!},\\
 Z_r^{(u_{\max},\infty)}
 &\le\frac{\Gamma(r+4,u_{\max})}
 {5040\pi^2\vartheta^4(r+3)!}.
\end{align}
Here $\Gamma(s,x)$ is the upper incomplete gamma function.
The middle integral is enclosed by complex ball quadrature~\cite{Johansson2017,Johansson2018}.
Ball arithmetic represents each real quantity by a midpoint and an error radius, propagated through the numerical operations.
For complex quadrature, we analytically continue $\sin^2\delta_3$ as $b^2/(a^2+b^2)$.

The calculation uses the exact inputs $\vartheta=1/100$ and $u_r=15/4$ or $23/4$.
Moments through $r=557$ are enclosed using 1024-bit arithmetic.
The relative and absolute integration tolerances for each energy subinterval are both set to $2^{-480}$.
The integration endpoints in $u=E/T$ are $u_{\min}=2^{-128}$ and $u_{\max}=2048$.
Both integral tails are included explicitly.
The trial functions contain up to 48 odd modes through degree 97, and their action is enclosed explicitly through degree 1023.
The high-order matrix elements are alternating sums of the energy moments.
Every explicitly computed matrix-element interval has a radius below $3\times10^{-15}$.
Equation~\eqref{eq:tail} bounds all odd rows from 1025 onward.
For the trial functions used in Table~\ref{tab:certificates}, this analytic tail contributes less than $3\times10^{-15}$ to the squared residual.

At each positive real frequency, the bound $\Theta_{j_0}(\phi)$ of the actual trial function is added to the explicitly computed finite-row residual.
Equation~\eqref{eq:residual} then gives a disk containing the full response.
This disk is centered at $\hat\kappa_\phi$ in the complex response plane, with radius given by the upper bound in Eq.~\eqref{eq:residual}.
Its projection onto the real axis gives an enclosure for $\Ree\hat\kappa$.
The trial coefficients have fixed finite binary real and imaginary parts obtained from numerical solutions.
Errors from solving the trial system are therefore included in the residual.
The center $\hat\kappa_\phi$ is itself enclosed by ball arithmetic.

The Galerkin trial solves the kinetic equation projected onto the retained Hermite modes.
Its full-equation residual is $\varepsilon_\phi$ in Eq.~\eqref{eq:kineticresidual}.
This projection does not in general minimize $\|\varepsilon_\phi\|$, so the residual norm need not decrease monotonically as the trial space grows.
Each trial gives an enclosure that independently contains the exact response.
For each state and frequency, Table~\ref{tab:certificates} uses the narrowest of these valid enclosures from trials with 16, 32, 40, and 48 modes.
Subtracting the real-response intervals gives Table~\ref{tab:certificates}.

\bibliographystyle{apsrev4-2}
%

\end{document}